\documentclass[9pt,twocolumn,twoside]{osajnl}
\journal{ol}
\setboolean{shortarticle}{true} 

\usepackage[percent]{overpic}
\usepackage{bm, upgreek}

\usepackage[overlay,absolute]{textpos}
\newcommand\PlaceText[3]{%
	\begin{textblock*}{10in}(#1,#2)
		#3
	\end{textblock*}
}%

\title{Dysprosium-doped ZBLAN fiber laser\\tunable from 2.8 $\upmu$m to 3.4 $\upmu$m, pumped at 1.7~$\upmu$m}

\author[1,*]{Matthew R. Majewski}
\author[1]{Robert I. Woodward}
\author[1]{Stuart D. Jackson}

\affil[1]{MQ Photonics, Department of Engineering, Faculty of Science and Engineering, Macquarie University, North Ryde, NSW 2109, Australia}
\affil[*]{Corresponding author: matthew.majewski@mq.edu.au}

\ociscodes{(140.3510) Lasers, fiber; (140.3070) Infrared and far-infrared lasers; (140.5680) Rare earth and transition metal solid-state lasers.}

\begin{abstract}
We demonstrate a mid-infrared dysprosium-doped fluoride fiber laser with a continuously tunable output range of 573~nm, pumped by a 1.7~$\upmu$m Raman fiber laser.
This represents the largest tuning range achieved to date from any rare-earth-doped fiber laser and critically, spans the 2.8--3.4~$\upmu$m spectral region which contains absorption resonances of many important functional groups and is uncovered by other rare-earth ions.
Output powers up to 170~mW are achieved, with 21\% slope efficiency.
We also discuss the relative merits of the 1.7~~$\upmu$m pump scheme, including possible pump excited-state absorption.

\end{abstract}

\setboolean{displaycopyright}{true}

\begin{document}

\maketitle

\PlaceText{25mm}{9mm}{Vol. 43, Issue 5, pp. 971-974 (2018); https://doi.org/10.1364/OL.43.000971}

Coherent light sources in the mid-infrared (mid-IR) are enabling many new spectroscopy and sensing technologies based on probing the strong, distinctive mid-IR absorption features of important functional groups and technical materials.
This opens up a range of applications across multiple disciplines such as gas sensing for defence and agricultural monitoring~\cite{kumar2012stand}, and breath analysis for medical diagnosis~\cite{wang2009breath}.
To practically exploit these opportunities, mid-IR lasers are required which cover a broad spectral range, while also being compact and rugged for deployment in resource-limited environments.

While there has been significant progress in the development of octave-spanning supercontinuum sources and frequency combs~\cite{Petersen2014mid,Hudson2017toward,Schliesser2012mid}, they remain very limited in spectral brightness, which limits detection sensitivity. 
It is therefore preferable for many applications to consider narrow-band tunable laser sources, offering orders-of-magnitude greater power spectral densities.
Indeed, widely wavelength tunable sources based on optical parametric amplification (OPA) processes have been recently demonstrated as ideal sources for molecular detection~\cite{Ruxton2012concentration,barria2014simultaneous,lambert2015broadband}.
However, OPA-based devices require complex, costly and highly sensitive optical arrangements, which limit their widespread potential.

\begin{figure}[htbp]
\centering
\includegraphics[width =0.9\linewidth]{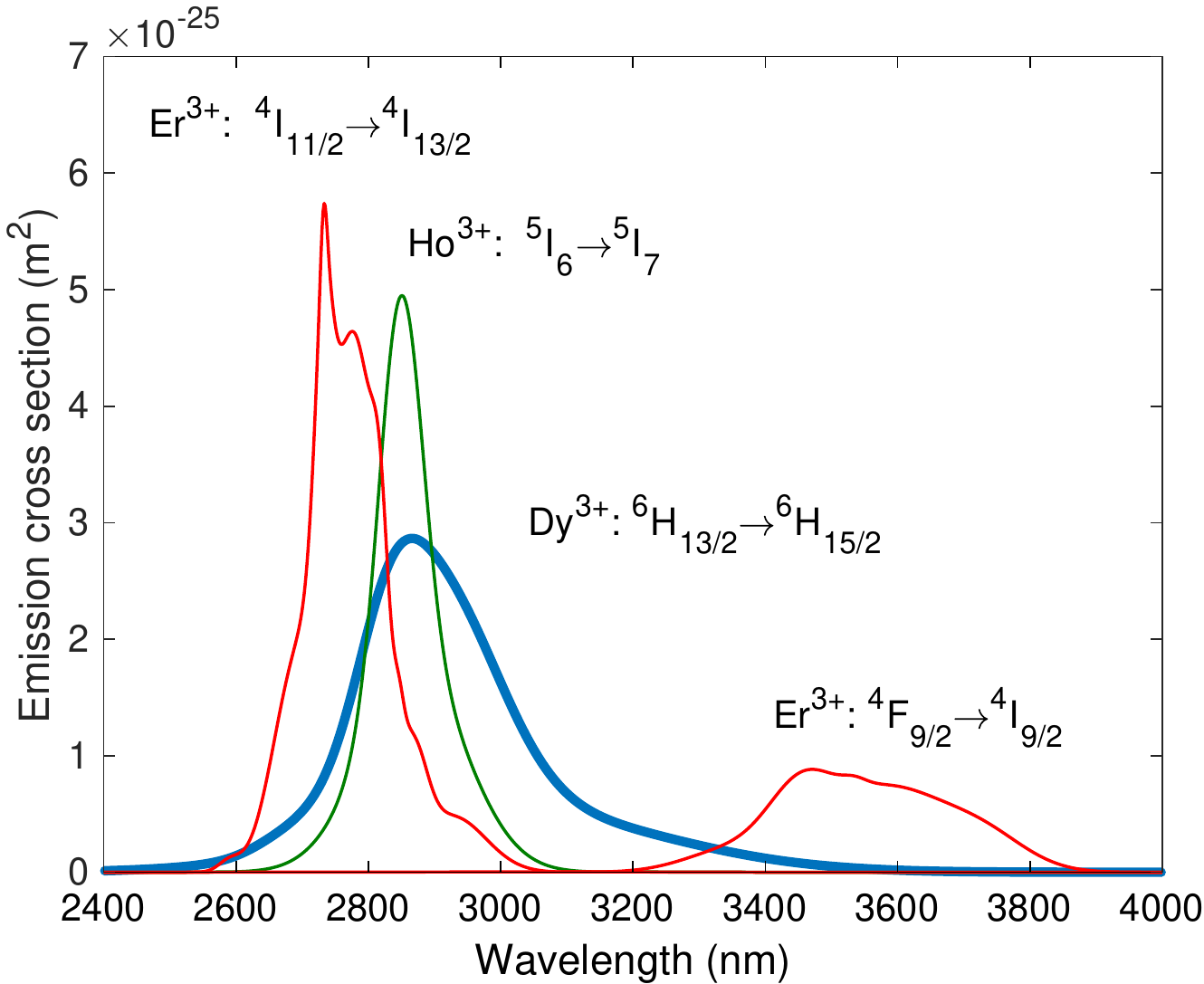}
\caption{Mid-infrared emission cross sections of rare-earths in ZBLAN. Measured data from Refs \cite{gomes2010energy,li2012numerical,henderson2016versatile}.}
\label{fig:erhody}
\end{figure}

An alternative route to high-power, tunable emission is rare-earth-doped fluoride fibre laser technology, which is emerging as a promising platform for compact mid-IR source development. 
The majority of mid-IR fiber lasers to date have employed erbium or holmium ions, which exhibit relatively broad emission cross sections in an amorphous fluoride glass host (shown in Fig.~\ref{fig:erhody}). 
Greater than 100~nm of tuning around 2.8~$\upmu$m has been demonstrated for both the  $^4I_{11/2}\rightarrow^4I_{13/2}$ transition of erbium (e.g. 2.71--2.83~$\upmu$m~\cite{libatique2000field}) and the $^5I_{6}\rightarrow^5I_{7}$ transition of holmium (e.g. 2.83--2.98~$\upmu$m~\cite{Crawford2015}). 
For longer wavelength emission, the $^4F_{9/2}\rightarrow^4I_{9/2}$ transition of erbium has recently been shown to be broadly tunable with a range of 450 nm centered around 3.5~$\upmu$m~\cite{henderson2016versatile}.
However, as Fig.~\ref{fig:erhody} reveals, a large spectral region around 3.2~$\upmu$m is currently unserved by rare-earth-doped lasers, yet is highly critical as it contains the absorption features of important functional groups, while also being in a atmospheric transmission window.

Bridging of this gap can be accomplished by the $^6H_{13/2}\rightarrow^6H_{15/2}$ transition of the relatively understudied dysprosium ion~\cite{jackson2003continuous, gomes2010energy}, which exhibits an emission cross section significantly broader than erbium and holmium (Fig.~\ref{fig:erhody}).
A recently demonstrated mid-IR dysprosium fiber laser confirmed this idea, exhibiting a 400~nm tuning range from 2.95 to  3.35~$\upmu$m~\cite{majewski2016tunable}.
This system was based on an in-band 2.8~$\upmu$m pump scheme~\cite{majewski2016highly}, however, and thus was unable to access the full gain bandwidth of the transition.
Previous reports of Dy:fluoride fiber lasers have also considered 1.1~$\upmu$m~\cite{jackson2003continuous} and 1.3~$\upmu$m~\cite{tsang2006efficient} excitation wavelengths, although both have been shown to suffer from strong pump excited-state absorption (ESA) which severely impacts performance~\cite{gomes2010energy}.
The optimum pump wavelength for mid-IR dysprosium-doped lasers thus remains an open question.

In this Letter, we demonstrate a dysprosium-doped ZBLAN fiber laser pumped at 1.7~$\upmu$m for the first time. 
This pump wavelength allows access to the entire gain bandwidth of the $^6H_{13/2}\rightarrow^6H_{15/2}$  transition, resulting in continuously tunable emission from 2.807 to 3.380~$\upmu$m, while also minimizing the quantum defect compared to shorter near-IR pump wavelengths. 
To our knowledge this is the widest tuning range yet achieved from a rare-earth-doped laser and also currently the only fiber laser source capable of targeting molecular absorption of OH and NH compounds around 2.9~$\upmu$m as well as important CH-based compounds such as methane which exhibit absorption around 3.4~$\upmu$m.

\begin{figure}
\centering
\begin{overpic}[width=0.9\linewidth]{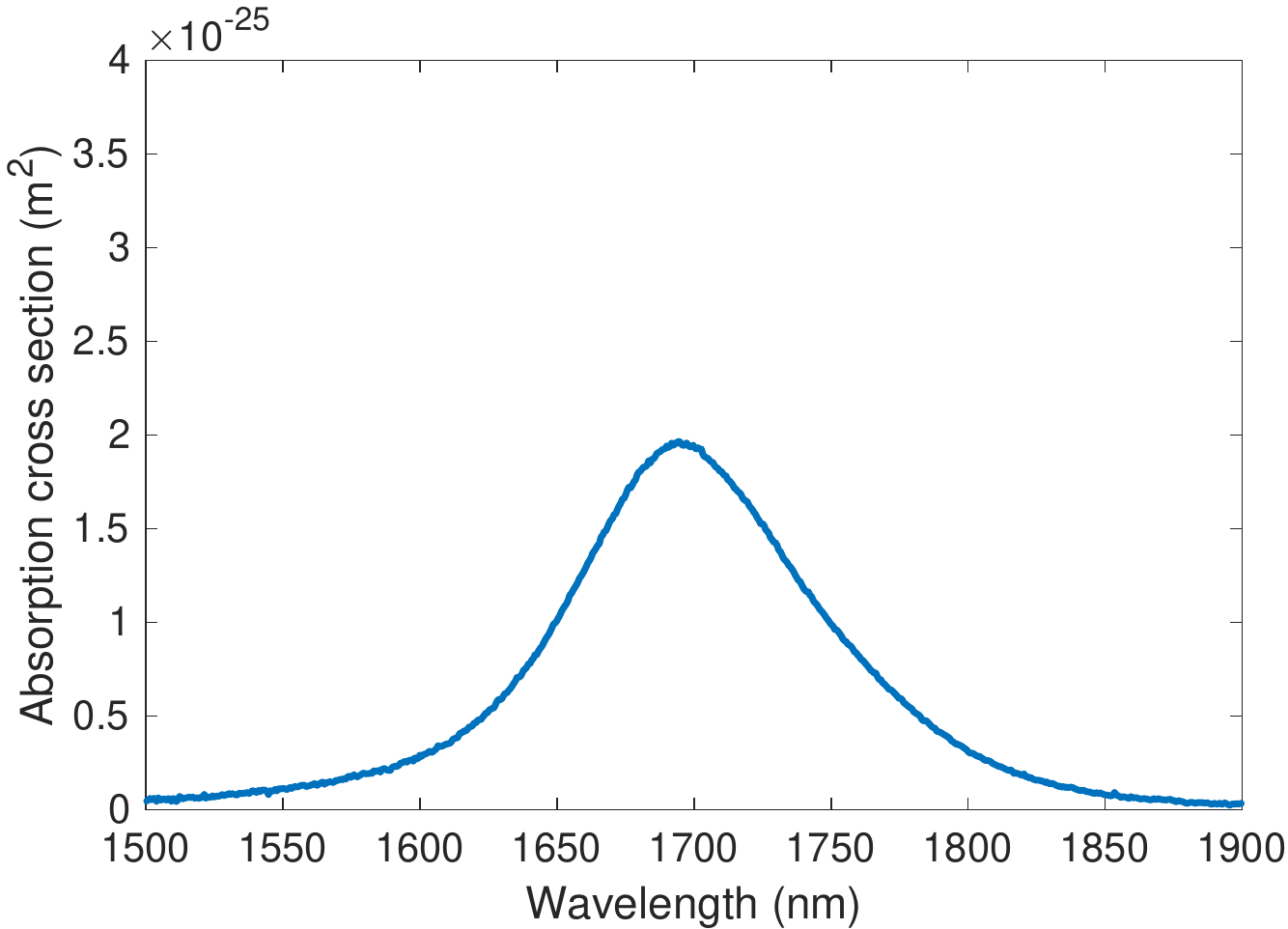}
	\put(57,35){\includegraphics[width=0.35\linewidth]{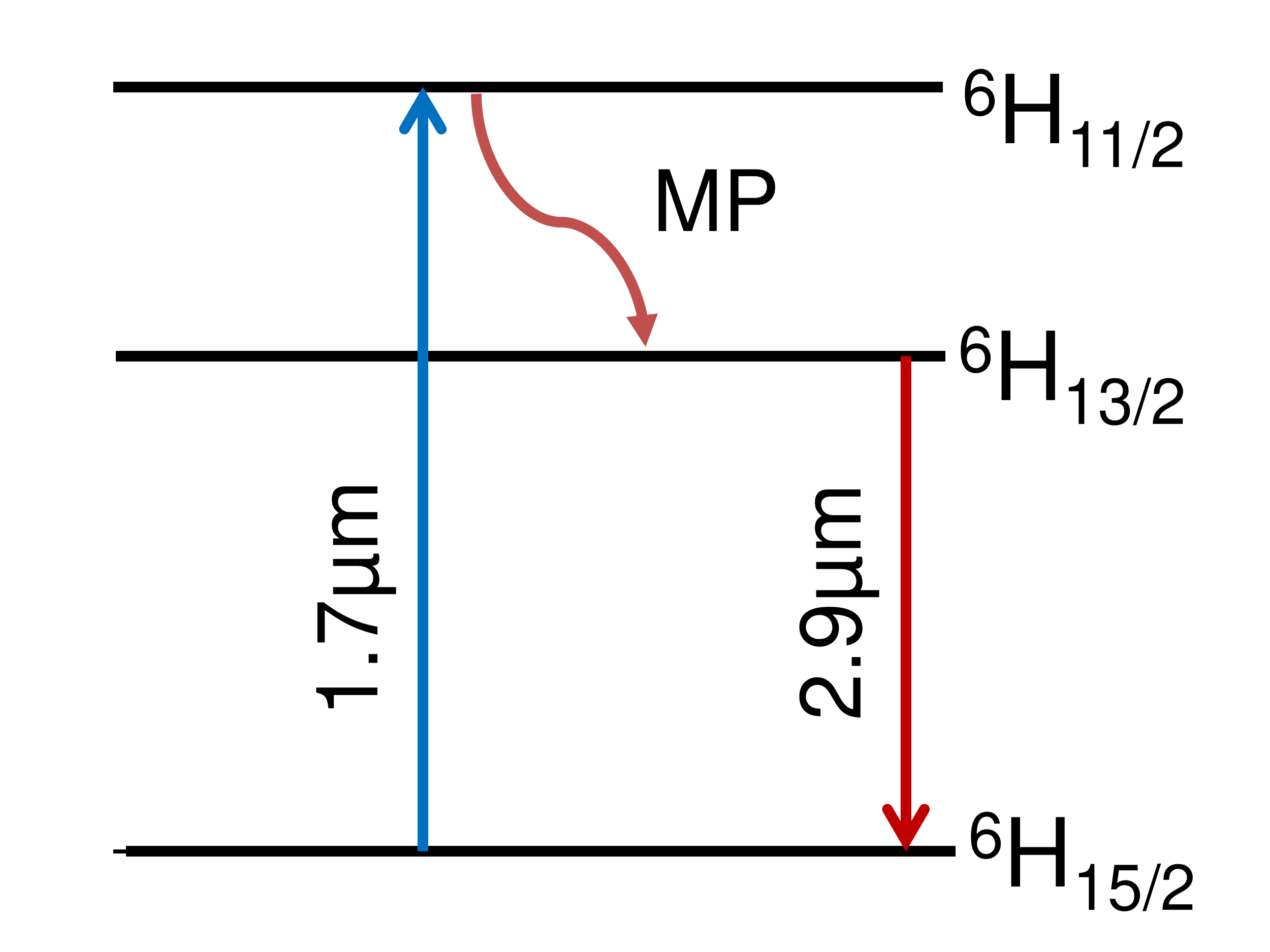}}

\end{overpic}
\caption{ Measured ground state absorption cross section of dysprosium in ZBLAN, centered around 1.7~$\upmu$m. The inset shows the simplified energy level diagram indicating target pump absorption, multiphonon (MP) relaxation, and 3$\upmu$m emission.}
\label{fig:energy}
\end{figure}

The relevant energy levels for this laser system are shown in the inset of Fig.~\ref{fig:energy}.
Lasing on the $^6H_{13/2}\rightarrow^6H_{15/2}$ ground-state transition when pumping the $^6H_{11/2}$ level is a quasi-three-level system.
The measured dysprosium absorption cross-section in a ZBLAN host for this pump scheme is shown in Fig.~\ref{fig:energy}, centered at $\sim$1.7~$\upmu$m.
As a ground state-terminated transition, generally a high level of inversion is required for efficient laser action, thus a high brightness (i.e. fiber laser) pump source is desired.

While commercial high-brightness sources around 1.7~$\upmu$m are not readily available, recently Raman fiber lasers (RFLs) have been shown to conveniently access this spectral region \cite{liu2014high}. We are able to produce the required pump power with a relatively simple in-house constructed RFL arrangement. 
Briefly, a diode-pumped Er/Yb co-doped fiber laser (Nufern) emitting around 1570 nm is coupled into 6~km of standard single-mode telecommunications fiber. 
Fiber Bragg grating (FBG) cavity mirrors (with 100\% and 10\% reflectivity) are spliced to this fiber, resonating the Stokes field and producing a narrow linewidth output centered at 1.7~$\upmu$m with up to 1.6~W continuous-wave (CW) power.

\begin{figure}[htbp]
\centering
\includegraphics[scale=0.28]{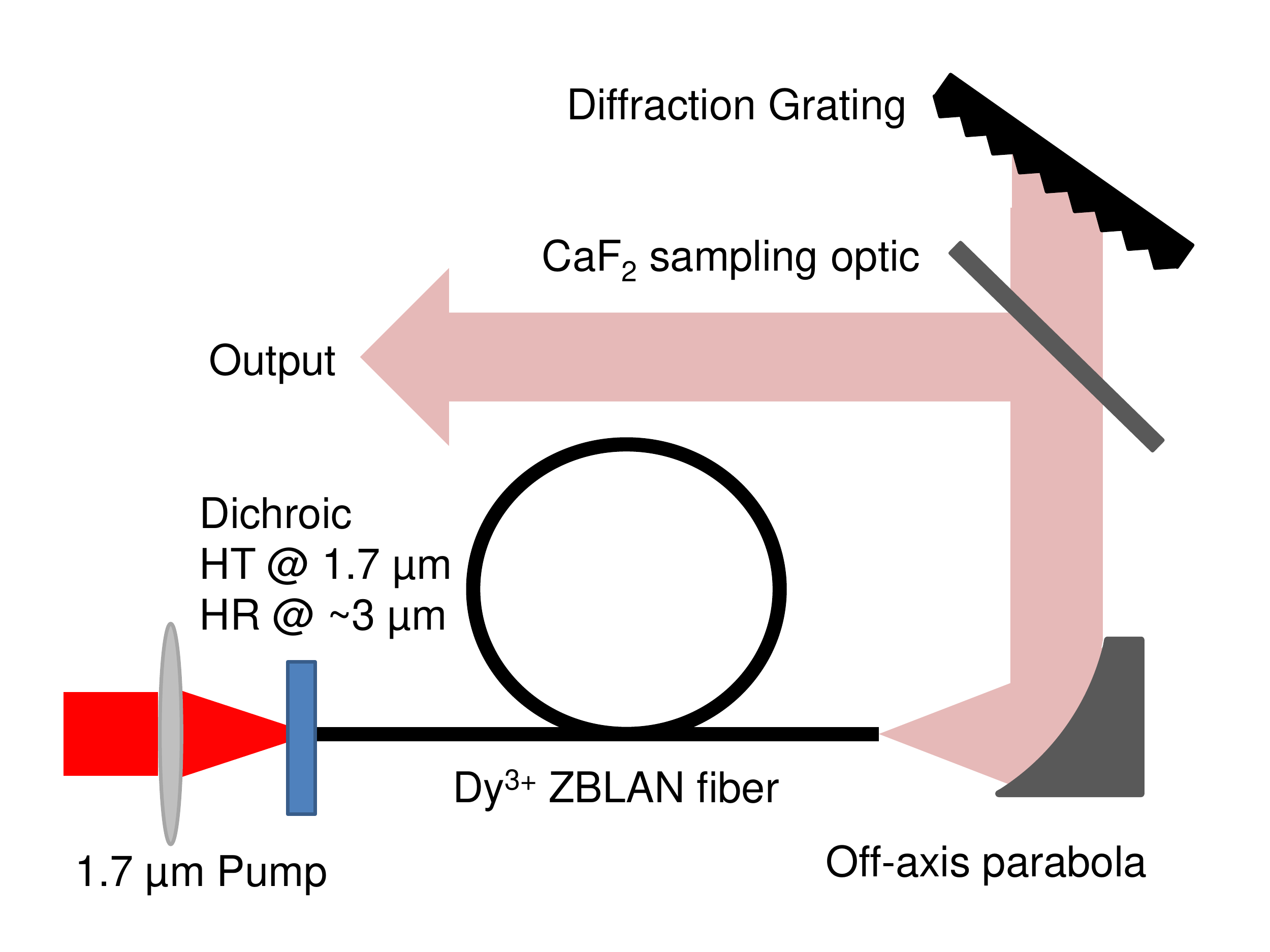}
\caption{Experimental setup for the tunable dysprosium fiber laser}
\label{fig:labsetup}
\end{figure}

The experimental setup for the tunable mid-IR laser is shown in Fig.~\ref{fig:labsetup}.
The dysprosium-doped ZBLAN fiber (Le Verre Fluor\'{e}) used in this work had a Dy$^{3+}$ concentration of 2000 ppm (3.63$\times10^{25}$ m$^{-3}$), a core diameter of $12~\upmu$m and a numerical aperture of 0.16, resulting in a single-mode cut-off of 2.6~$\upmu$m. 
Though the fiber was few-moded at the pump wavelength, efficient coupling into the fundamental mode was achieved by collimation and focusing of the RFL output with matched $f=25$~mm anti-reflection coated aspheric lenses, resulting in a coupling efficiency of nominally 85\%.
A dichroic mirror transmitting pump light and broadband highly reflective around 3~$\upmu$m is butt-coupled to the fiber input. 
Output is collimated with an off-axis parabolic reflector and a diffraction grating in the Littrow configuration closes the cavity. 
The parabolic reflector was chosen over a comparable focal length lens to ensure that no focal position shift occurred over the entire range of laser operation. 
An uncoatced CaF$_2$ window is placed at 45 degrees between the parabolic reflector and the grating to sample laser output.

\begin{figure}[htbp]
\centering
\includegraphics[width=\linewidth]{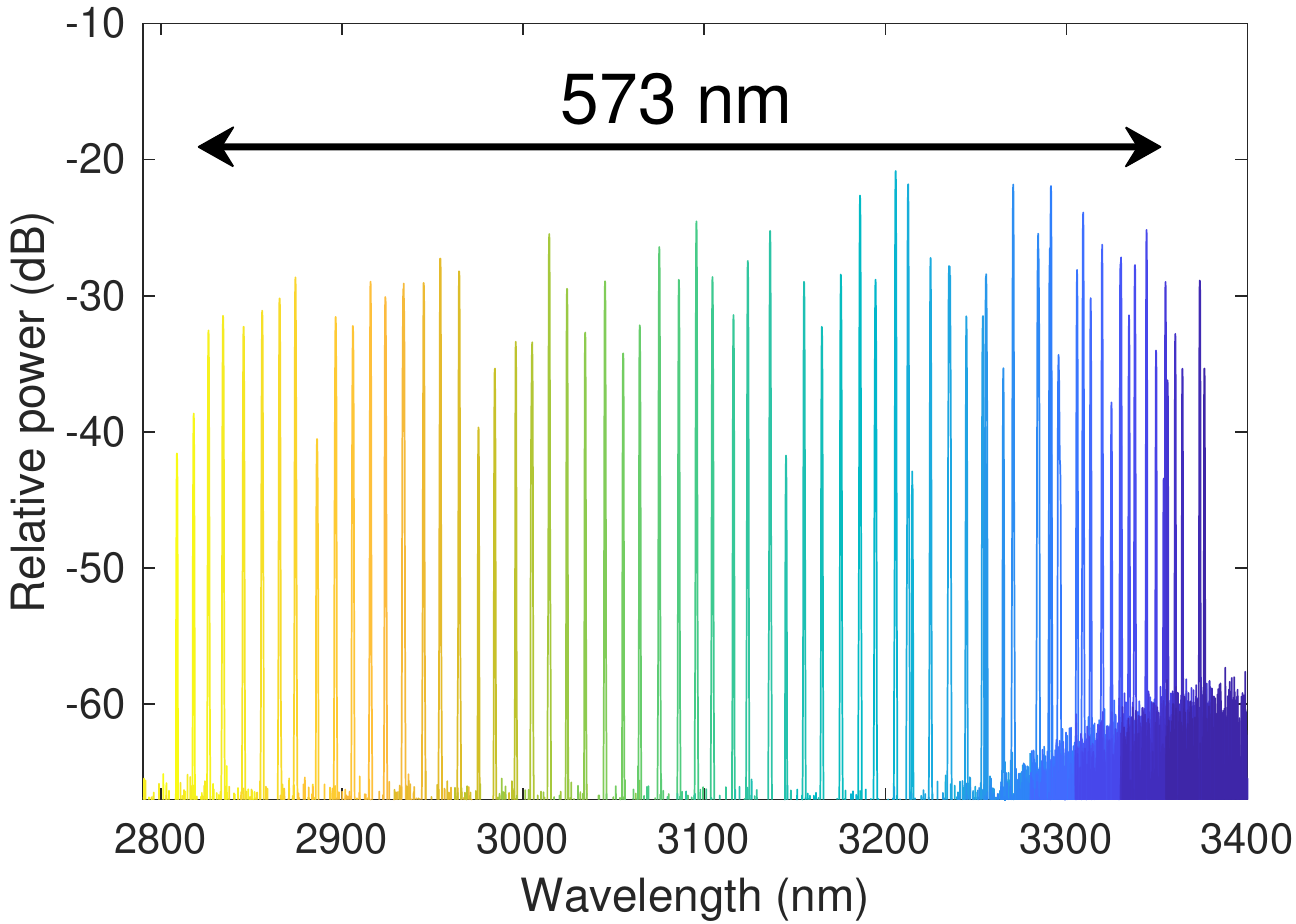}
\caption{Measured optical spectra over the laser emission tuning range}
\label{fig:tuning}
\end{figure}

Initially, we explore the potential tuning range of the mid-IR laser output by operating the pump laser at its maximum 1.6~W output power.  
A 60 cm~length of active fiber is chosen so that $\sim$95\% of injected pump light is absorbed.
The spectrum of the sampled output is monitored with an optical spectrum analyzer (Yokagawa) as the angle of the diffraction grating is changed.The resultant optical spectra seen in Fig.~\ref{fig:tuning} reveal narrow linewidth laser emission which is continuously tunable   in the range 2.807 to 3.380~$\upmu$m, equating to almost 600 nm tunability.

 As seen in Fig.~\ref{fig:erhody}, the dysprosium emission cross section becomes comparatively quite small beyond 3.2$\upmu$m, thus requiring increasingly long gain lengths to compensate for cavity losses and reach laser oscillation. Despite the small cross section magnitude, the clear observation of laser emission in this long wavelength region suggest that further infrared tuning should  be possible with an increased fiber length.
 However, for our current system, it is found empirically that further increase in fiber length increases the oscillation threshold beyond our pump power limitation. 
 At present, however, the near-600~nm tunability represents the broadest spectral tunability to date from a rare-earth-doped laser: an extension of 150~nm compared to the previously broadest result using erbium:ZBLAN fibre centered at 3.5~$\upmu$m~\cite{henderson2016versatile}. 

\begin{figure}[htbp]
\centering
\includegraphics[scale=0.6]{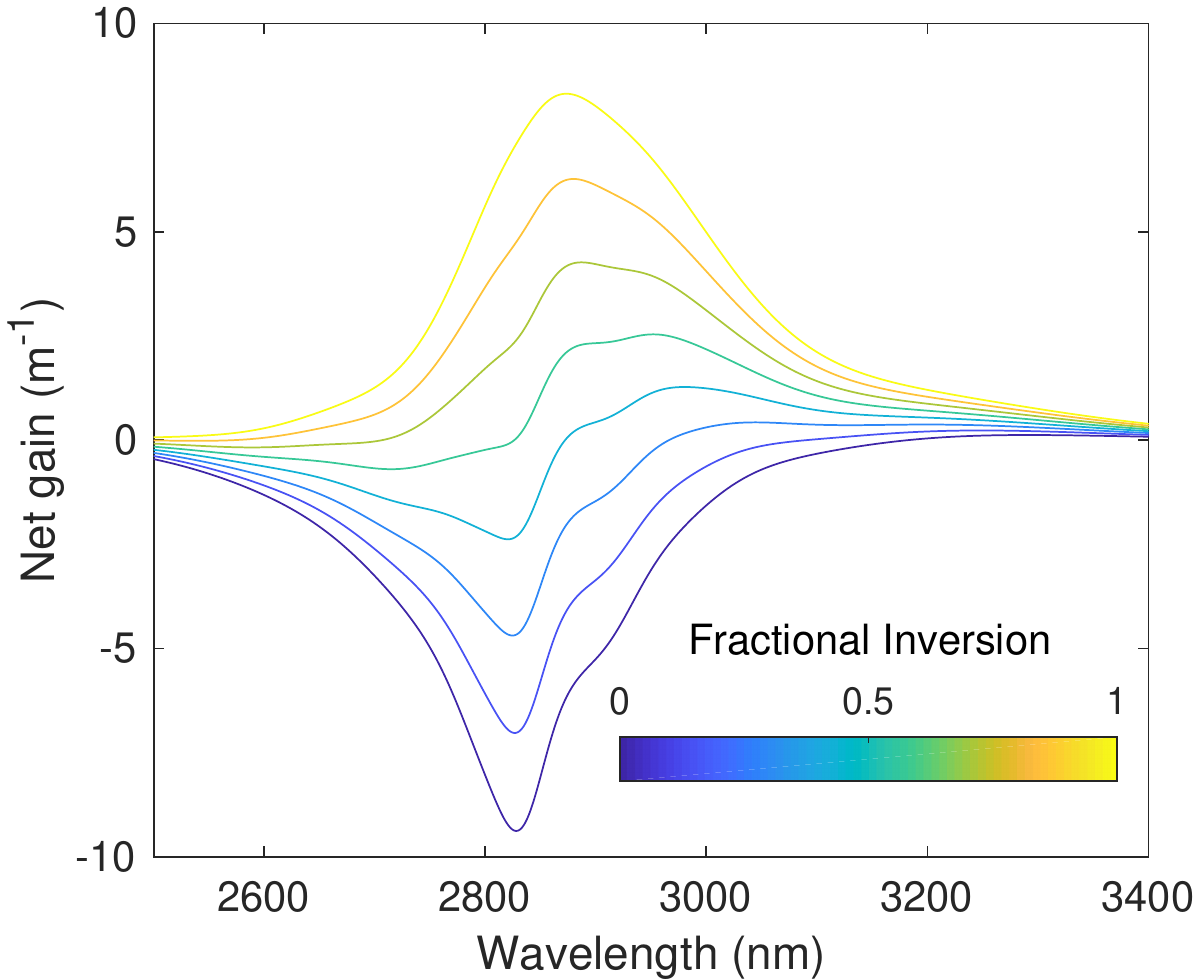}
\caption{Net gain  ($g_{\mathrm{net}}=\sigma_eN_2-\sigma_aN_1$) calculated from measured cross sections of dysprosium in ZBLAN for increasing fractional population inversion }
\label{fig:inversion}
\end{figure}

The short wavelength range of this system represents a 150~nm increase  over the previous demonstration of a tunable dysprosium fiber laser \cite{majewski2016tunable}. This is a direct result of the pump wavelength chosen here, as the previous system was based on an inband pumping scheme, setting a fundamental lower limit on emission wavelength.
Further tuning to shorter wavelengths is limited by signal re-absorption and the fairly strong overlap of absorption and emission cross sections that dysprosium exhibits.
 While the emission cross sections are still substantial below 2.8~$\upmu$m, Fig.~\ref{fig:inversion} shows that net-positive gain ($g_{\mathrm{net}}=\sigma_eN_2-\sigma_aN_1$, where $\sigma_{e,a}$ are emission and absorption cross sections respectively and $N_{1,2}$ are the relevant level populations) below 2.8~$\upmu$m is only possible for high levels of population inversion. 

\begin{figure}[htbp]
\centering
\includegraphics[scale=0.6]{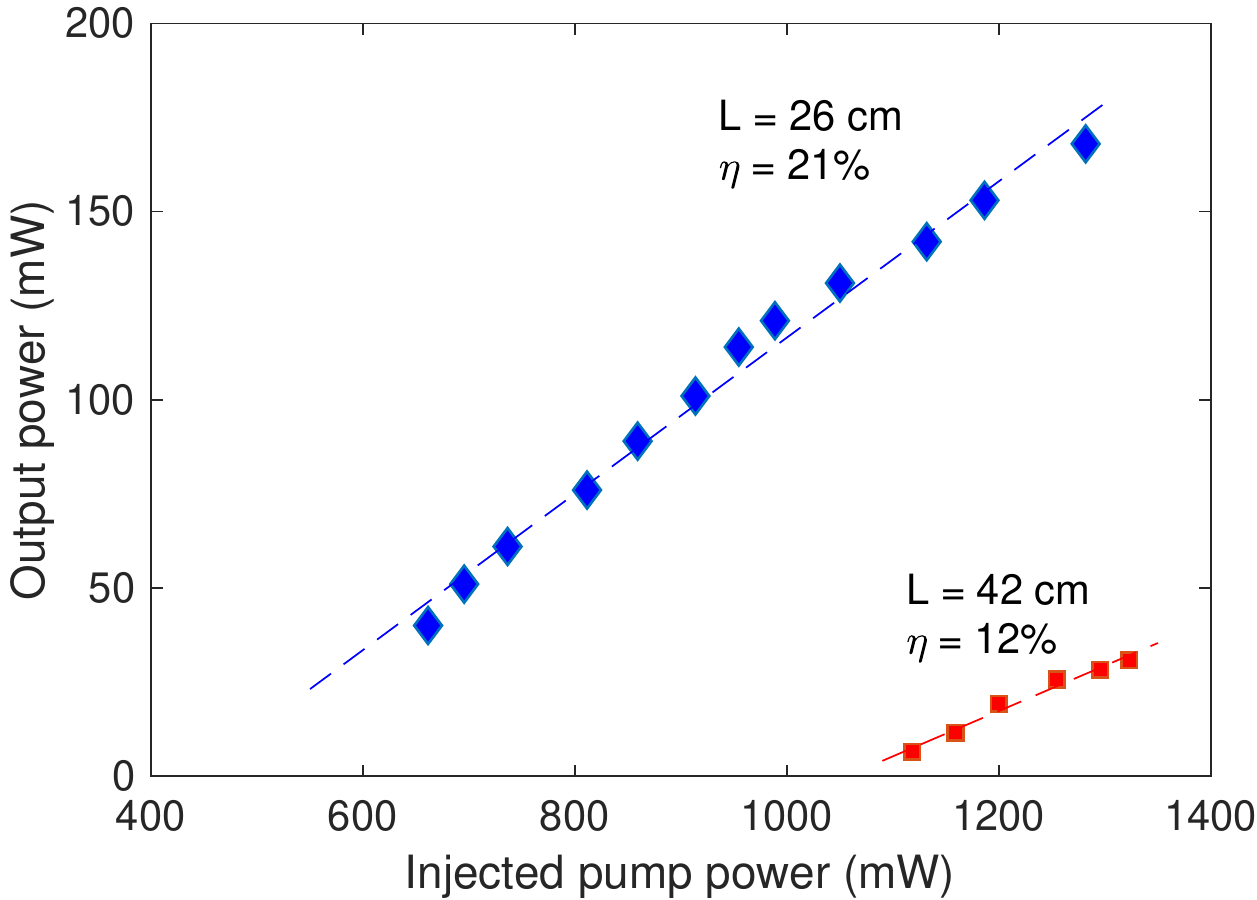}
\caption{Free-running laser output power as a function of injected pump power for two lengths of dysprosium fiber. L=26~cm produces a slope efficiency ($\eta$) of 21\% and a maximum output power of 170mW }
\label{fig:slope}
\end{figure}

Having demonstrated the wideband tuning potential, we now focus on optimization of the laser for high power and efficiency, and consider the relative merits of the 1.7~$\upmu$m pump transition.
We replace the parabolic reflector and diffraction grating arrangement with a butt-coupled output dichroic mirror which is highly reflective at the pump wavelength and nominally 50\% reflective across the 3~$\upmu$m band. 
This results in free-running laser emission around the gain peak of 2.95~$\upmu$m.
The output power as a function of injected pump power is presented in Fig.~\ref{fig:slope}. 
For a 42~cm length of fiber, the slope efficiency is found to be 12\% with an oscillation threshold of 1.05~W. 
In an effort to reduce threshold and increase efficiency we note that this length of fiber exhibits nominally 90\% pump absorption, thus a substantially shorter length of fiber combined with pump retroreflection should be advantageous (similar to bidirectional pumping).
Thus, we utilize a 26~cm length of active fiber and measure a slope efficiency of 21\% and an injected pump power threshold of 440 mW. 
A maximum mid-IR output power of 170~mW is achieved, limited only by the available pump power.

\begin{figure}[htbp]
\centering
\includegraphics[scale=0.5]{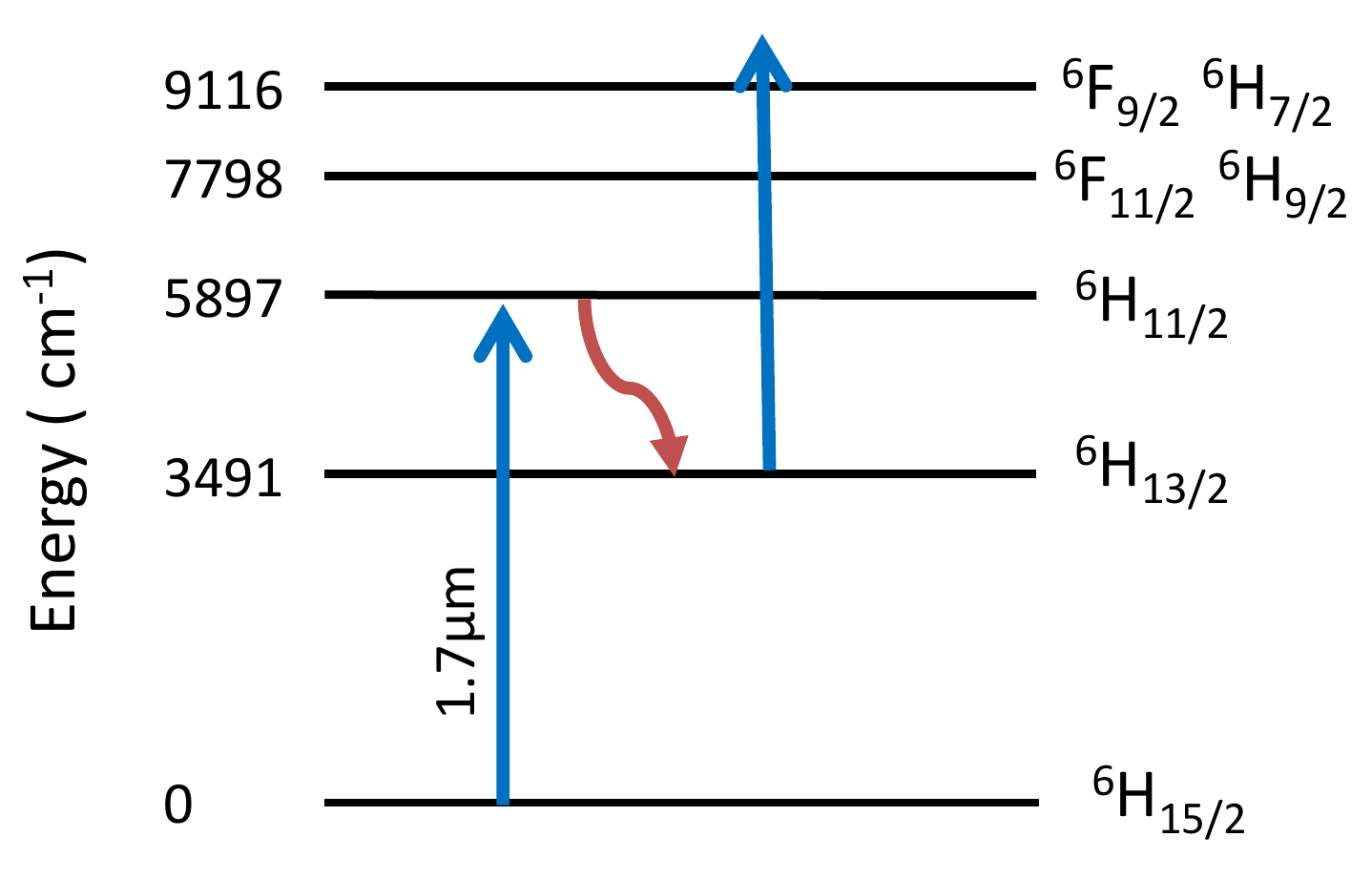}
\caption{Energy level diagram of dysprosium in ZBLAN showing the possible excited state absorption (ESA) of 1.7~$\upmu$m pump photons from the upper 3~$\upmu$m laser level  (energy level positions taken from \cite{adam1988optical})}
\label{fig:ESA}
\end{figure}

Though the maximum slope efficiency demonstrated for this system exceeds that of previous demonstrations of near-IR (1.1 \& 1.3 ~$\upmu$m) pumped mid-IR dysprosium fiber lasers \cite{jackson2003continuous,tsang2006efficient}, it still falls short of the Stokes limit of 55\%. 
Both previous investigations attributed efficiency reductions principally to the detrimental influence of pump excited state absorption (ESA) originating from the upper laser level $^6H_{13/2}$, reducing population inversion. 
ESA is a possible explanation here as well; absorption of a 1.7~$\upmu$m pump photon from the $^6H_{13/2}$ level at 3491~cm$^{-1}$ results in a total energy of 9373~cm$^{-1}$ which is nearly resonant with the thermally coupled $^6F_{9/2}$ and $^6H_{7/2}$ levels near 9116~cm$^{-1}$ (see Fig.~\ref{fig:ESA}) . 
This suggestion is also supported by a recent observation of 1.7~$\upmu$m ESA from the $^6H_{13/2}$ level in a study on dysprosium-doped PGS crystals~\cite{jelinkova2013dysprosium}.
Empirically we note that the single pass pump absorption is well approximated by small signal absorption even at elevated injected power. 
As the saturation power ($ P_{sat}=Ah\nu/\sigma_a\tau $ where $A$ is core area, $h\nu$ is pump photon energy, $\sigma_a$ is absorption cross section, and $\tau$ is the upper state lifetime) here is calculated to be of the order of 100 mW, substantial saturation of pump absorption would be expected. 
Preliminary modeling of pump absorption based on rate equations, including the effect of ESA as a free parameter, produces an estimate of ESA cross section at 1.7~$\upmu$m on the order of 1$\times10^{-25}$ m$^{-2}$. 
Therefore, while the 1.7~$\upmu$m pump scheme appears promising in terms of the reduced quantum defect compared to shorter wavelength near-IR excitation, further investigation is required to accurately characterize the influence of ESA on this system.

\begin{figure}[b]
\centering
\includegraphics[scale=0.7]{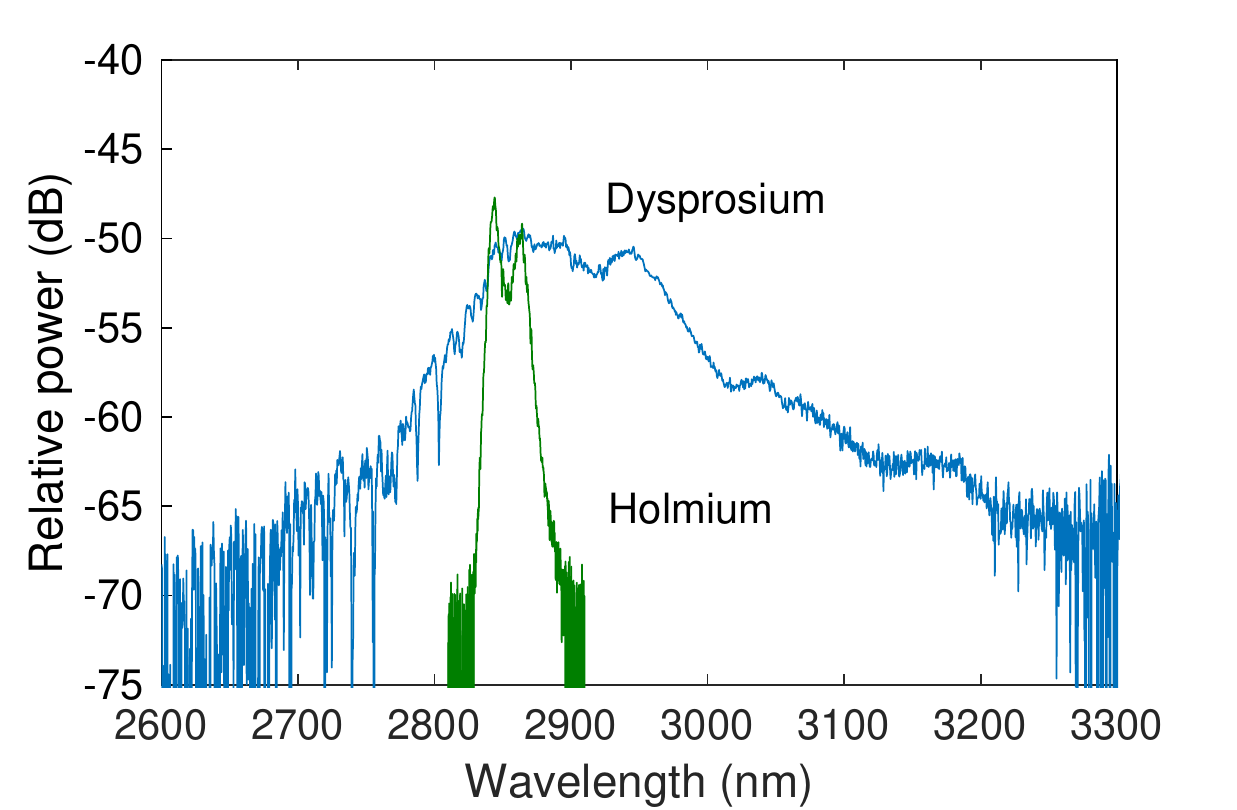}
\caption{Measured amplified spontaneous emission (ASE) spectrum of the dysprosium ZBLAN fiber. The measured ASE spectrum of holmium is also shown for comparison}
\label{fig:ASE}
\end{figure}

Beyond the promise of application as a widely tunable continuous wave laser source, the large gain bandwidth of this system also has strong implications for the possibility of mode-locked ultrafast operation; as the transform-limited pulse duration is inversely proportional to pulse bandwidth. 
Recent demonstrations of mid-IR ultrafast fiber lasers have opened up the possibility for a wide range of applications, many of which could benefit by increasingly shorter pulses and corresponding higher peak powers. 
Demonstration of an erbium-based fiber system reached the femtosecond level \cite{duval2015femtosecond} but the spectral bandwidth was reduced from the full gain bandwidth largely due to atmospheric water absorption which exhibits several strong features between 2.7 and 2.8~$\upmu$m. 
Further reduction in pulse duration to sub 200 fs was achieved with a holmium system with emission centered at a longer wavelength (away from deleterious water vapor absorption lines) of 2850 nm \cite{antipov2016high}; here the spectral width of the pulse closely approached the full gain bandwidth.
While shorter 70~fs pulses have been achieved through extracavity nonlinear compression~\cite{woodward2017generation}, simplicity and compactness advantages would result from the generation of ultra-broadband pulses directly from a mode-locked mid-IR oscillator. 
As a means of comparing our dysprosium fiber to these systems we measure the amplified spontaneous emission (ASE) spectrum (Fig.~\ref{fig:ASE}). 
As compared to holmium, dysprosium is substantially broader, with the bulk of the width even further removed from detrimental atmospheric absorption. 
Such spectral width implies sub-100 fs few-cycle pulses could be possible even without accessing the entirety of the gain bandwidth.

In conclusion, we have demonstrated the first mid-IR dysprosium-doped fiber laser pumped in the near-infrared at 1.7~$\upmu$m. 
With a simple diffraction grating based cavity we were able to achieve CW laser emission continuously tunable over a range of nearly 600 nm---this is both the widest tuning range yet achieved in any rare earth fiber laser and a substantial increase over previous results. 
Emission over this range covers absorption features of both OH/NH and CH functional groups which is not possible with current alternative rare-earth-doped sources. 
With an optimized fiber length, the maximum efficiency achieved for this pumping scheme exceeds that of previous near-infrared pumped dysprosium fiber lasers and we also highlighted the possibility of detrimental ESA at this pump wavelength.
Further extension of the tuning range and maximum output power should be possible with a higher power pump source and the implementation of bi-directional pumping.

Funding: Australian Research Council (ARC) (DP140101336).

We thank Professor Graham Town for providing the 6~km telecommunications fiber and Dr. Laercio Gomes for dysprosium spectroscopic data. RIW acknowledges support through an MQ Research Fellowship.

\end{document}